# Correlation functions quantify super-resolution images and estimate apparent clustering due to over-counting.

Sarah Veatch,*‡ Benjamin Machta,† Sarah Shelby,‡ Ethan Chiang,‡ David Holowka,‡ and Barbara Baird‡

*Department of Biophysics, University of Michigan, Ann Arbor, MI; Departments of †Physics and ‡Chemistry and Chemical Biology, Cornell University, Ithaca, NY

## Abstract:

We present an analytical method to quantify clustering in super-resolution localization images of static surfaces in two dimensions. The method also describes how over-counting of labeled molecules contributes to apparent self-clustering and how the effective lateral resolution of an image can be determined. This treatment applies to clustering of proteins and lipids in membranes, where there is significant interest in using super-resolution localization techniques to probe membrane heterogeneity. When images are quantified using pair correlation functions, the magnitude of apparent clustering due to over-counting will vary inversely with the surface density of labeled molecules and does not depend on the number of times an average molecule is counted. Over-counting does not yield apparent co-clustering in double label experiments when pair cross-correlation functions are measured. We apply our analytical method to quantify the distribution of the IgE receptor (FcεRI) on the plasma membranes of chemically fixed RBL-2H3 mast cells from images acquired using stochastic optical reconstruction microscopy (STORM) and scanning electron microscopy (SEM). We find that apparent clustering of labeled IgE bound to FcεRI detected with both methods arises from over-counting of individual complexes. Thus our results indicate that these receptors are randomly distributed within the resolution and sensitivity limits of these experiments.

Address reprint requests and inquiries to sveatch@umich.edu

# Introduction

Recent advances in super-resolution imaging have enabled imaging of cellular structures at close to molecular length-scales using light microscopy [1,2,3,4]. In conventional fluorescence microscopy, the average distance between fluorescently labeled molecules is typically small compared to the point spread function (PSF) of the microscope (~250nm). In this limit, the detailed fluorescence character of individual labeled molecules does not contribute significantly to the final image, since many individual labeled molecules are averaged within the PSF of the measurement. Super-resolution imaging and localization techniques can improve lateral resolution by an order of magnitude or more (typically ~20nm). In this limit, the average distance between neighboring labeled molecules is frequently close to the resolution of the measurement, and the fluorescence properties of individual labeled molecules can significantly impact the resulting images. For example, under-sampling of super-resolution images can lead to lower effective resolution by some measures, as discussed in previous work [5].

This manuscript addresses issues that arise from over-counting signals from individual labeled molecules in super-resolution images of two dimensional surfaces. Over-counting occurs when molecules are labeled with more than one primary or secondary antibodies, when labeling antibodies are conjugated to multiple fluorophores, when multiple signals are detected across the PSF, or when the same fluorophore is counted two or more times because it cycles reversibly between an activated and dark state. In all of these cases, over-counting leads to the appearance of self-clustering over distances that correspond to the effective PSF of the measurement. In this manuscript we first describe a method to quantify the distribution of labeled molecules in images and develop a simple model to predict the magnitude of apparent clustering arising from over-counting. For cases where over-counting is deliberate, we can apply this formalism to measure the effective lateral resolution of an image reconstructed from localized signals. Our analytical approach is then applied to high resolution images of the high affinity IgE receptor (FcεRI) on the surface of RBL-2H3 mast cells obtained using scanning electron microscopy (SEM) or by stochastic optical reconstruction microscopy (STORM). Our approach can also be applied to other types of high resolution imaging methods such as transmission electron microscopy (TEM), photoactivated light microscopy (PALM/fPALM), Stimulated Emission Depletion (STED) microscopy, and near field optical scanning microscopy (NSOM).

## Results and Discussion

**Pair correlation functions quantify over-counting**

Pair correlation functions quantify organization in heterogeneous systems, and are easily applied to super-resolution localization data. The pair auto-correlation function, $g(r)$, reports the probability of finding a second localized signal a distance r away from a given localized signal, is efficiently calculated using fast Fourier transforms, and can account for complex boundary shapes without additional assumptions. Detailed methods used to calculate correlation functions are described in Materials and Methods, and a Matlab function to calculate $g(r)$ from images is supplied in Supplementary material.

If an ensemble of molecules is distributed on a two dimensional surface with centers at positions $\vec{r}$ described by the density function $\rho(\vec{r})$ and an average density $\langle \rho(\vec{r}) \rangle = \rho$, the associated pair auto-correlation function of molecular centers is:

$$g(\vec{r}) = <\rho(\vec{R})\rho(\vec{R}-\vec{r})>/\rho^2,$$

where the average is over all positions $\vec{R}$ in the image. Often it is assumed that $g(\vec{r})$ is symmetric to rotations, and it is averaged over angles to obtain $g(r)$. At $r=0$, $g(r)$ contains a delta function, $\delta(r)$, with magnitude of $1/\rho$. Correlation functions are plotted for $r > 0$ since $g(r=0)$ is a trivial contribution. However, if $g(r)$ is calculated from an image obtained from a measurement with finite resolution in the presence of over-counting, the measured correlation function will contain a remnant of this delta function at nonzero radius:

$$g_{meas}(r) = [\delta(r)/\rho + g(r>0)] * g_{psf}(r)$$

Where $g_{psf}(r)$ is the correlation function of the PSF of the measurement, $g(r=0)$ represents the correlation function for the distribution of labeled molecules, and $*$ denotes a convolution. The convolution acts to smear $\delta(r)$ to finite radius. If we assume a Gaussian-shaped form of the PSF with a half width of σ, this becomes:

$$g_{meas}(r) = \exp\{-r^2/4\sigma^2\}/\pi\sigma^2\rho + g(r>0)*g_{psf}(r) \quad \text{(Eqn 1)}$$

The first term of $g_{meas}(r)$ arises from over-counting of labeled molecules with finite resolution and is inversely proportional to the average density of labeled molecules (ρ). The second term describes the distribution of labeled molecules within the resolution limits imposed by the PSF, and is independent of the density of labeled molecules. If we interpret the PSF as the probability of detecting a signal at a given position with respect to the center of the labeled molecule, and we assume that the detected signals sample the PSF of each molecule through a random process according to this probability, then this equation holds independently of the average number of times each molecule is counted (derivation in Materials and Methods). This is graphically depicted in Figure 1 for the example of labeled molecules partitioned either randomly or into circular domains. In the special case of a random distribution of labeled molecules, $g(r>0)=1$ and

$$g_{meas}(r) = g_{psf}(r)/\rho + 1$$
$$= \exp\{-r^2/4\sigma^2\}/\pi\sigma^2\rho + 1 \quad \text{(Eqn 2)}$$

Another commonly used methodology to quantify heterogeneity in membrane systems is the modified Ripley's K function, denoted (L(r)-r)/r. L(r) is related to the average number of signals within a radius r of a given particle, which is the integral of $2\pi r g(r)$ [6]. As a result, Ripley's methods are not well suited to quantify images that are subject to over-counting, since the integration propagates apparent clustering due to over-counting to long distances. By contrast, the correlation function is not much affected by over-counting at distances larger than the width of the PSF. This is demonstrated in Figure 1D and 1E and a detailed relationship between g(r) and (L(r)-r)/r is presented in Materials and Methods.

The estimates of apparent-clustering due to over-counting presented in the first terms of Eqns. 1 and 2 are valid only when over-counting occurs via a random process. This is expected to be the case for the majority of high-resolution measurements that are subject to over-counting, such as stochastic blinking of fluorophores in STORM measurements and reversible switching of fluorescent proteins in some PALM measurements. This case should also apply when over-counting occurs through conjugation of multiple organic fluorophores to proteins or ligands, or when labeling of proteins with primary and secondary antibodies. Eqns. 1 and 2 are also expected to apply when high-resolution images are obtained without using localization imaging techniques, such as STED or NSOM. In these experiments, a large number of photons map out the extended PSF of the measurement explicitly. In all of these examples, the average surface density of labeled molecules can be extracted from measured correlation functions as long as labeled molecules are randomly distributed. This feature of measured correlation functions has been used previously to determine the oligomerization state of labeled molecules in diffraction limited images [7].

Our estimates of clustering will not be accurate if over-counting is not randomly distributed over all labeled molecules. The first terms of Eqns. 1 and 2 will over-estimate apparent clustering from over-counting for cases where labeled molecules are counted less frequently than expected from a random distribution. This would apply, for example, when detection of a signal from a labeled molecule decreases the probability that the same labeled molecule will be detected additional times. This occurs in a STORM or PALM measurements if there is a significant probability of bleaching a fluorophore after it is activated. If imaging is conducted in a manner that ensures that all labeled molecules are counted at most once, then measured correlations are due only to clustering of labeled molecules. This would be the case if every activated fluorophore is irreversibly bleached after being counted in a PALM/fPALM measurement, or if a labeling strategy was employed that ensured only single gold particles label target proteins in EM measurements.

The first terms of Eqns. 1 and 2 will underestimate the magnitude of apparent clustering when labeled molecules are over-counted more frequently than expected from a random distribution. This would occur, for example, when the act of counting a signal from a labeled molecule increases the probability that additional signals will be detected from the same labeled molecule. This condition occurs in STORM or PALM/fPALM measurements if activated probes are counted once for each frame for which they are imaged, including the same signal remaining activated for multiple observations. In cases where probes are deliberately over-counted, it is possible to isolate the over-counting term in Eqn. 1 and directly measure the effective PSF of the measurement. This is accomplished by tabulating correlation functions from two images reconstructed from the same set of acquired signals. The first image is reconstructed from intentionally over-counted signals (e.g. where signals localized in the same position in sequential frames are counted independently), while the second image is reconstructed from signals where over-counting is avoided (e.g. when localized signals correlated in time and space are grouped together). Subtracting $g_{meas}(r)$ of the grouped image from $g_{meas}(r)$ of the intentionally over-counted image results in a curve that is proportional to $g_{psf}(r)$, since the second term of Eqn. 1 is independent of the number of times a labeled molecule is counted. In an ideal experiment, the range of $g_{psf}(r)$ will be simply related to the localization precision of the average signal. In most cases, this calculated $g_{psf}(r)$ will be broader than the localization precision due to contributions from other experimental limitations such as stage drift and mobility of labeled molecules [8]. This is approach is applied to determining the apparent lateral resolution of a reconstructed STORM image in Figure 2.

### Pair correlation functions quantify heterogeneity

For cases when measured correlation functions contain contributions that cannot be attributed to over-counting, such as when $g_{meas}(r) >> 1$ for $r >> \sigma$ then the residual correlations can be attributed to clustering of labeled molecules. Much information can be extracted to discern the underlying structural distribution by monitoring both the shape and the magnitude of the correlation function. For example, the average number of labeled molecules per cluster is given by $\langle NC \rangle = 1 + \rho \int_0^\infty (g(r)-1) 2\pi r \, dr$ and the effective potential of mean force (PMF) between labeled molecules is given by $PMF(r) = -k_B T \ln\{g(r)\}$ [9]. The shape of the correlation function also sheds light on the physical basis that governs heterogeneity [10]. Three examples of different simulated particle distributions are shown in Figure 3A, their calculated correlation functions shown in Figure 3B have distinct features that can be used to distinguish the organizing principles giving rise to these distributions. Simulations of particles placed within a series of circular domains produce correlation functions that are damped oscillations, where the frequency of the oscillation corresponds to the average domain size, and the decay length quantifies correlations between neighboring domains [11]. By contrast, simulations of particles distributed in fluctuations produce correlation functions that decay as exponentials [12]. Both micro-emulsion (circle) and fluctuation models have been proposed as physical mechanisms that could produce small and subtle heterogeneity in resting cell plasma membranes [13,14], and in principle the shape of correlation functions can be used to distinguish these different models.

### Examples of over-counting in super-resolution data

We apply this quantitation to two types of super-resolution data obtained with labeled IgE specifically bound to the high affinity FcεRI receptor on RBL-2H3 mast cells. Figure 4A shows a STORM image of Alexa-647 fluorophores conjugated directly to IgE on the bottom surface of a chemically fixed cell. In STORM measurements, the majority of probes are forced into a reversible dark state in the presence of bright light, a reducing environment, and basic pH [4]. This enables imaging and localization of a sparse subset of fluorophores. Probes stochastically switch between bright and dark states and high resolution images can be reconstructed when samples are imaged over time, as described in Materials and Methods.

Correlation functions derived from cells labeled with IgE show significant auto-correlations at short distances, as shown in Figure 4B. We estimate the average surface density of IgE-FcεRI to be 200 complexes/μm$^2$ based on previous measurements [15]. Using this density along with the measured effective PSF, we find that the size and shape of the measured correlation function is well described by Eqn 2. Residual correlations are small, consistent with the conclusion that labeled FcεRI-IgE complexes are randomly distributed within the membrane.

This analytical approach can also be applied to scanning electron microscopy (SEM) images where target proteins are labeled with primary antibodies followed by secondary antibodies conjugated to gold particles as described in Materials and Methods. Figure 5 shows a flat section of the top surface of a RBL-2H3 cell with FcεRI-IgE complexes that are immuno-labeled with 10nm gold particles. This labeling scheme allows for multiple gold particles to decorate individual target proteins, and the correlation function detects clustering over short distances (Figure 5B). In this experiment, the PSF is governed by the finite size of labeling antibodies and not by the precision of localizing the gold particles. The magnitude of measured correlations is well described by Eqn. 2 using 200 complexes/$\mu m^2$ as the surface density of IgE-FcεRI.

Direct evidence that apparent clustering of labeled IgE-FcεRI complexes arises from over-counting is provided by double-label SEM experiments, where distinguishable but functionally identical pools of IgE-FcεRI are labeled with differently sized gold particles. Similar to auto-correlation, the cross-correlation function, c(r), quantifies the increased probability of finding a distinguishable particle a distance r away from a given particle of a different type. Unlike the auto-correlation function, the cross-correlation function does not contain a delta function at r=0, and is therefore it is not affected by over-counting even when an experiment is conducted with finite resolution. The experiment was conducted by first creating two separate pools of FcεRI on the cell surface by pre-incubating the cells with a mixture of IgE labeled with either the fluorophore Alexa488 or the fluorophore FITC prior to fixation. These were distinctively labeled with fluorophore-specific primary antibodies of different species followed by species-specific secondary antibodies conjugated to gold particles of different sizes (Figure 5C). By this scheme, small and large gold particles cannot bind to the same FcεRI protein because only a single IgE antibody binds to each FcεRI protein (Mendoza and Metzger, 1976)

If IgE-FcεRI complexes are clustered at the cell surface, then cross-correlation functions between the two functionally identical pools of IgE will show significant co-clustering. In contrast, we find that cross-correlation functions (black points in Figure 5D) indicate random distributions within experimental error bounds given by counting statistics. This comparison shows that the appearance of clustering in single label images is dominated by over counting single target proteins.

These observations contradict several previous studies which report that IgE-FcεRI complexes are pre-clustered in unstimulated RBL-2H3 cells [16,17,18]. Since similar strategies were used to label IgE-FcεRI in these studies, we expect that apparent self-clustering from over-counting was incorrectly identified as self-clustering of labeled proteins. It is possible that previous reports of self-clustering of other membrane components visualized by electron microscopy can also be attributed to over-counting, since labeling schemes oten require the use of multiple or polyclonal antibodies. This potential pitfall of electron microscopy labeling and imaging was noted in early work that contributed to the Fluid Mosiac model of membranes [19].

In conclusion, we demonstrate that correlation functions provide an analytical tool to quantify heterogeneous distributions of labeled molecules in super-resolution experiments in the presence of signal over-counting that gives rise to the appearance of short-range clustering. We present an analytical method that predicts the magnitude of correlations arising from over-counting, and describe a method to measure the apparent PSF of an image for cases when signals can be intentionally over-counted. We have applied this analysis methodology to quantifying the lateral distribution of IgE-FcεRI complexes on the surface of unstimulated RBL-2H3 cells imaged using STORM and SEM. We find that IgE-FcεRI complexes are randomly distributed by both methods after considering expected contributions from over-counting, and this interpretation is confirmed by a direct measurement of cross-correlation functions in double label SEM measurements. These examples emphasize the importance of explicitly considering over-counting when quantifying images of proteins in membranes, where the extent of heterogeneity may be small and subtle.

## Materials and Methods

### Chemicals and Reagents

Rabbit anti-Alexafluor 488, reactive Alexafluor 647 and 488 were purchased from from Invitrogen (Eugene, OR). Mouse anti-FITC, 10nm gold-conjugated anti-rabbit IgG (whole molecule), 10nm gold-conjugated anti-mouse IgG (whole molecule), 5nm gold-conjugated anti-rabbit IgG (whole molecule), β-mercaptoethanol, Glucose Oxidase, and Catalase were purchased from Sigma (St. Louis, MO). 5nm gold-conjugated anti-mouse was purchased from GE Healthcare (Piscataway, NJ). A488-IgE and FITC-IgE were prepared by conjugating purified mouse monoclonal anti-2,4-dinitrophenyl (DNP) IgE with Alexafluor 488, Alexafluor 647, or FITC (Invitrogen) as previously described [20,21]. Glutaraldehyde (25% stock) was purchased from Ted Pella (Redding, CA). Para-formaldeyde was purchased from Electron Microscopy Services (Hatfield, PA).

### Stochastic optical reconstruction microscopy (STORM)

*Sample preparation*: Rat Basophilic Leukemia (RBL-2H3) cells were cultured as described previously [20], then harvested using Trypsin-EDTA, and plated sparsely overnight at 37ºC in glass-bottom MatTek dishes (Ashland, MA). The cells were sensitized with A647-labeled IgE (1µg/ml) in media for 1 to 2 hours at room temperature. Dishes containing cells were rinsed with additional media, and following a 5 minute incubation at 37ºC, rinsed again with PBS before fixation (4% paraformaldehyde 0.1% glutaraldehyde in PBS) for 10 minutes at room temperature. Samples were then blocked with 2% fish gelatin, 2 mg/mL BSA in PBS for 10 minutes.

*Imaging:* Samples were imaged on an inverted microscope (Leica DM-IRB, Wetzlar, Germany) under through-objective TIRF illumination by a 100mW 642nm diode laser (Crystalaser, Reno, NV). Images were captured with an Andor iXon 897 EM-CCD camera (Belfast, UK) using custom image acquisition code written in Matlab (Mathworks, Natick, MA). To induce A647 photo-switching, cells were imaged in the presence of an oxygen-scavenging and reducing buffer containing 100mM Tris, 10mM NaCl, 10% w/w glucose, 10ug/mL glucose-oxidase, 2ug/mL catalase, and 1% β-mercaptoethanol at pH 8.4. Movies of A647 photo-switching were acquired at 25 frames per second for ~2500 frames and analyzed by localizing the centers of diffraction limited spots through least squares fitting using the *fminfunc()* function in Matlab. Localized centers were culled to exclude outliers in brightness, width, aspect ratio, and localization precision. Centers were combined when the same fluorophores were in sequential frames if the localization position was found within a circle of radius 2x the maximum localization precision of the population. Reconstructed images are assembled from the remaining centers and convolved with a Gaussian for display purposes.

**Scanning Electron Microscopy (SEM)**

*Sample Preparation:* RBL-2H3 mast cells were grown overnight to ~50% confluency on 2 mm x 2 mm silicon chips at 37˚C under standard cell culture conditions [22], and high affinity IgE receptors (FcεRI) were labeled with either A488-IgE (1µg/mL) (for single label experiments) or a 1:1 mixture of A488-IgE and FITC-IgE (total 1µg/mL) (for double label experiments) for 2-3 hr prior to the experiment. Cells were washed quickly in phosphate buffered saline (PBS), and immediately fixed in 4% (w/v) p-formaldehyde and 0.1% (w/v) glutaraldehyde for 10 min at room temperature in PBS. Fixed cell samples were washed in blocking solution (2 mg/mL BSA and 2% (v/v) fish gelatin in PBS) and labeled sequentially with primary antibodies and gold conjugated secondary antibodies in the presence (usual) or absence (control) of 0.1% Triton X-100 in blocking solution. Incubations were 1 h at room temperature with wash steps in between. After labeling, the cell samples were further fixed in 4% p-formaldehyde and 1% glutaraldehyde for 5 min at room temperature, and then thoroughly washed in distilled water. Following dehydration through a series of graded ethanol washing steps, samples were critical point dried, mounted on round aluminum SEM stubs, and sputtered with carbon to prevent charging. For single label experiments the primary antibody was rabbit anti-Alexafluor 488 and the 10nm gold conjugated secondary antibody was goat anti-rabbit IgG.  For double label experiments, the primary antibodies were mouse anti-FITC and rabbit anti-Alexafluor 488, while the secondary antibodies were 5nm gold-conjugated anti-rabbit IgG and 10nm gold-conjugated anti-mouse IgG. Samples were labeled first with 10nm and then 5nm gold antibody conjugates.

*Imaging*:  Mounted samples were imaged with a Schottky field emission Scanning Electron Microscope (LEO 1550) at 20KeV. The dorsal (top) surfaces of intact, adherent cells were imaged using secondary electron detection (SED) and backscattered detection (BSD) at high magnification. Flat membrane regions were selected for imaging. For imaging 10 nm gold particles, individual micrographs were obtained at 35K magnification, and typical images cover 2.4 μm$^2$ of the cell surface. For imaging 5 nm gold particles and in double-label experiments with 10 and 5 nm gold particles, micrographs were obtained at 75K-100K magnification. Immuno-gold labeled protein distributions for >20 different cells and >3 individual experiments were obtained for all experimental conditions presented.   Gold particle centers were localized by finding the weighted centroid of identified particles using automated image processing software written in Matlab.  Reconstructed images are formed by convolving an image of the particle centers with a Gaussian shape with half-width given by the gold particle radius.

**Calculation of Correlation Functions**

Pair auto- and cross- correlation functions were tabulated in Matlab using Fast Fourier Transforms (FFTs) as follows:

$$g(\vec{r}) = \frac{FFT^{-1}(|FFT(I)|^2)}{\rho^2 N(\vec{r})}$$

Where $FFT^{-1}$ is an inverse Fast Fourier Transform, I is a binary image where pixels have a value of 1 at localized centers and all other pixels have a value of 0, and $N(\vec{r})$ is a normalization that accounts for the finite size of the acquired image.  The image I is padded with zeros in both directions out to a distance larger than the range of the desired correlation function (maximally the size of the original image) to avoid artifacts due to the periodic nature of FFT functions.  The normalization factor N is the autocorrelation of a window function W that has the value of 1 inside the measurement area, and is also padded by an equal number of zeros.

$$N(\vec{r}) = FFT^{-1}(|FFT(W)|^2)$$

This normalization is essentially the total squared area over which the correlation function is calculated accounting for the fact that there fewer possible pairs separated by large distances due to the finite image size. When calculating correlation functions from STORM images, the cell interior was first masked, and this mask was then used as the window function W.

Cross-correlation functions between SEM images of 5nm and 10nm particle centers ($I_{5nm}$ and $I_{10nm}$ respectively) are tabulated in a similar manner:

$$c(\vec{r}) = \text{Re}\left\{ \frac{FFT^{-1}(FFT(I_{5nm}) \times conj[FFT(I_{5nm})])}{\rho_{5nm}\rho_{10nm}N(\vec{r})} \right\}$$

Here $conj[]$ indicates a complex conjugate, $\rho_{5nm}$ and $\rho_{10nm}$ are the average surface densities of 5nm and 10nm particles respectively, and Re{} indicates the real part. This computation method of tabulating pair auto and cross-correlations is mathematically identical to brute force averaging methods. Correlation functions were angularly averaged by first converting to polar coordinates using the Matlab command *cart2pol()*, and then binning by radius. $g(r)$ values are obtained by averaging $g(\vec{r})$ values that correspond to the assigned bins in radius. Errors in $g(r)$ are standard errors of the mean.

**Calculation of Modified Ripley's K Functions**

The statistical significance of clustering can also be determined using the Ripley's K function, which measures the increased density of particles within a circle of radius r and is related to the pair correlation function through integration:

$$K(r) = \int_0^r g(r')2\pi r' dr'$$

Frequently, Ripley's K function is restated when plotting the results from electron microscopy studies [23]:

$$L(r) - r = \sqrt{K(r)/\pi} - r = \sqrt{2\int_0^r g(r')r' dr'} - r$$

Furthermore, $L(r) - r$ curves reported in the literature are typically normalized to a confidence interval, so that the amplitudes of normalized $L(r) - r$ traces indicate the statistical significance of clustering within a radius r. Confidence intervals of $L(r) - r$ are calculated by propagating the statistical errors of $g(r)$ through $L(r) - r$. Errors in $g(r)$ are dominated by counting statistics that vary inversely with the square root of the number of particles found at a given distance.

**Detailed Derivation of Equations 1 and 2**

Consider a set of molecules at positions $\vec{r}_i$ for $1 < i < N$ with density $\rho = N/A$ and A the total area. The actual density of molecules as a function of $\vec{r}$ is given by $\rho(\vec{r}) = \sum_i \delta(\vec{r} - \vec{r}_i)$. The correlation function of these molecules is given by (normalized to 1 in the case where there are no correlations, as in previous sections):

$$g(\vec{r}) = \frac{1}{A\rho^2} \int d\vec{R} \rho(\vec{R}) \rho(\vec{R} + \vec{r}) = \frac{1}{AN^2} \sum_{i,j} \delta(\vec{r}_i - \vec{r}_j - \vec{r}) = \frac{1}{\rho}\delta(\vec{r}) + g(\vec{r} > 0)$$

Where in the last step we have defined $g(\vec{r} > 0)$ as the correlation function with the terms where $i = j$ removed. Consider stochastically building this correlation function by repeated measurements of individual emitters. We assume that every measurement uniformly and independently probes a random one of the N emitters. We further assume that a probe that actually resides at point $\vec{r}$ will be measured at $r'$ with probability given by $P(\vec{r}'|r) = PSF(\vec{r}' - \vec{r})$ with $PSF(\vec{r})$ the normalized point spread function. In that case each measurement is independently distributed according to the observed distribution $\rho_{meas}(\vec{r}) = \frac{1}{N}\sum_i PSF(\vec{r} - \vec{r}_i)$. After making M independent measurements of particle positions $r'_i$ the density of measurements is $\rho_{meas} = M/A$ and we can construct a correlation function:

$$g_{meas}(\vec{r}) = \frac{A}{M^2} \sum_{k,l} \delta(\vec{r}_k' - \vec{r}_l' - \vec{r})$$

We can relate our observed correlation function, $g_{meas}(\vec{r})$, to the actual $g(\vec{r})$ as follows. Firstly there is a term coming from those terms wher $k = l$, which, as it is always at exactly at $\vec{r} = 0$ we can easily ignore. When $k \neq l$ then each of $\vec{r}_k'$ are independently taken from the distribution given in the equation above. As such, we expect to measure not the bare $g(\vec{r})$ but instead:

$$\langle g_{meas}(\vec{r}) \rangle = \frac{1}{\rho_{meas}}\delta(\vec{r}) + \frac{A}{N^2}\sum_{i \neq j}\int d\vec{R} PSF(\vec{R}) PSF(\vec{r}_i - \vec{r}_j - \vec{R} - \vec{r})$$

$$= \frac{1}{\rho_{meas}}\delta(\vec{r}) + \frac{1}{\rho}g_{PSF}(\vec{r}) + g_{PSF}(\vec{r}) * g(r > 0)$$

Where the star denotes a convolution, and $g_{PSF}(\vec{r})$ denotes the correlation function of the point spread function with itself. Notice that the density of measurements only enters into the first term which we can easily throw out, since it is always measured at $\vec{r} = 0$. The result presented here is only valid in the regime where each measurement of a particle position is nearly independent. Additional correlations could result from deviations from a Poisson distribution in the number of times a single emitter is counted, which, as discussed in detail in previous sections is common, though not universally the case.

## Acknowledgements

We thank Prabuddha Sengupta and Amit Singhai for helpful conversations and assistance with experiments.  Research was funded through the NIH (SLV: K99GM087810; BAB: R01AI18306).

## Cited References

# Figure Legends

**Figure 1: Simulated demonstration of apparent clustering arising from over-counting individual labeled molecules with a finite PSF.** (A) Labeled molecules centered at black stars are convolved by a Gaussian PSF with half-width σ (red areas). Blue points are examples of signals detected with probability given by the intensity of the red area. Here the over counting ratio (OCR) is 3, meaning each labeled molecule is counted on average 3 times. (B) Red labeled molecules are confined within gray circular domains with 25 pixel radius, while green labeled molecules are distributed at random. Both labeled molecules have an average surface density $\rho = 2\times10^{-3} / \text{pixel}^2$ and $\sigma = 2 \text{ pixels}$. (C) Correlation functions calculated from B for structures as indicated. Red (green) signals are sampled at random from red (green) PSF areas with OCR=1, as described in A. g(r) for red centers and gray domains are equivalent within error, but g(r) for red signals shows additional clustering at short r, in agreement with Eqn 1. Green signals are also clustered at short r as described by Eqn 2, while g(r) for green centers is random within error. (D) Simulated g(r) for labeled red molecules partitioned into the gray domains as in B but with different average surface densities (ρ). Apparent clustering at short r decreases as ρ is increased, but long range correlations are unchanged, consistent with Eqn 1. (E) Modified Ripley's functions, (L(r)-r)/r, calculated from clustered red centers is slightly lower than but resembles functions calculated for red signals at large r. As expected, modified Ripley's functions for randomly distributed green centers do not show significant clustering over any radius. In contrast, functions calculated from green signals show significant apparent clustering over large distances.

**Figure 2: Measuring effective resolution of reconstructed super-resolution images with explicit over-counting.** (A,B) Reconstructed STORM images of labeled IgE on the bottom surface of RBL-2H3 mast cells. The region enclosed in the red box is magnified in the right panel. The image shown in A is reconstructed from raw data where each localized signal is counted independently. In B, explicit over-counting arising from probes remaining activated for multiple frames is removed by grouping localized signals that are strongly correlated in both time and space. Grouping methods are described in Materials and Methods. (C) Correlation functions are measured from both the raw image to obtain $g_{raw}(r)$ and from the grouped image to obtain $g_{group}(r)$. The correlation function of the raw image contains more apparent clustering at short radii than the measured correlation function of the grouped image because there are additional contributions in the raw image from over-counting. Subtracting $g_{group}(r)$ from $g_{raw}(r)$ results in a curve that is proportional to the correlation function of the effective point spread

function of the measurement, $g_{PSF}(r)$. In this example, the black points are fit assuming a Gaussian PSF, $g_{PSF}(r) = A\exp\{-r^2/4\sigma^2\}$, where σ is determined to be 9.5nm and A is an constant related to the number of times each probe was deliberately over-counted.

**Figure 3: Correlation functions quantify heterogeneity.** A) Simulated particle distributions are created by placing particles with radii of two pixels at random on pre-made templates. Three examples are shown: Small circles have radii between 4 and 8 pixels (left), large circles have radii between 10 and 30 pixels (center), and fluctuations are produced by simulating an Ising model at T = 1.075 $T_c$ (right), where $T_c$ is the critical temperature and the predicted correlation length (ξ) is ~4 pixel widths [12]. The top and bottom panels under each heading in A display the same particle distributions, while the bottom panels in A show both the particles and the template for demonstration purposes. Correlation functions are tabulated from a large number of simulations resembling the ones shown in the top panels (A). The correlation functions in B are fit to two different functional forms to account for distinct features in the curves. g(r) for the two circle distributions have a well defined dip below g(r)=1, and are fit to a damped cosine function: g(r) = 1+A×exp(-r/α)×cos(πr/2$r_o$), where A is an amplitude, α is a measure of the coherence length between circles, and $r_o$ is the average radii of clusters. This is the predicted functional form for a correlation function of a micro-emulsion [11]. The correlation function to the fluctuation model does not dip below g(r)=1 and is fit to the predicted form for critical systems: g(r) = 1+A×$r^{-1/4}$×exp(-r/ξ). From this example, it is apparent that both the shape and range of the correlation function reveal significant information regarding the underlying structure that gives rise to the heterogeneity. Also, when correlation functions are fit to the appropriate model, they accurately reproduce the radii of the circle distributions and the correlation length of the fluctuating distribution shown in part A.

**Figure 4: Apparent clustering of IgE-FcεRI observed by STORM is well described as over-counting of multiple signals from individual labeled molecules.** (A) STORM images of a representative fixed RBL-2H3 cell fixed after labeling with IgE directly conjugated to Alexa 647. Magnification of square inset shown at right. Localize centers are convolved with a Gaussian PSF with 50nm (whole cell) or 15nm (inset) half-width for display purposes. (B) Correlation functions of localized single molecule centers averaged over 8 cells closely resemble the prediction of Eqn 2 using σ=15±2.5nm and ρ=200±25/μm². Error bars on green points are propagated through $g_{PSF}/\rho$ which at short distances is the dominant source of error.

**Figure 5: Apparent clustering of IgE-FcεRI observed by SEM is consistent with over-counting of multiple signals from individual labeled molecules.** (A) A reconstructed image showing gold particles labeling IgE-FcεRI complexes on the top surface of a fixed RBL-2H3 cell. IgE-FcεRI is labeled post fixation with primary and gold-tagged secondary antibodies. (B) g(r) of localized gold particle centers is well described by Eqn 2 using σ=9±.5nm and ρ=200±50/μm$^2$. Error bars on green points are propagated through $g_{PSF}/\rho$. (C) 10nm and 5nm gold particles label distinct populations of IgE-FcεRI in double label experiments. (D) Auto-correlation functions of pooled particles quantify clustering in good agreement with Eqn 2. Cross-correlation functions, c(r), are not affected by over-counting and show no evidence for IgE self-clustering.

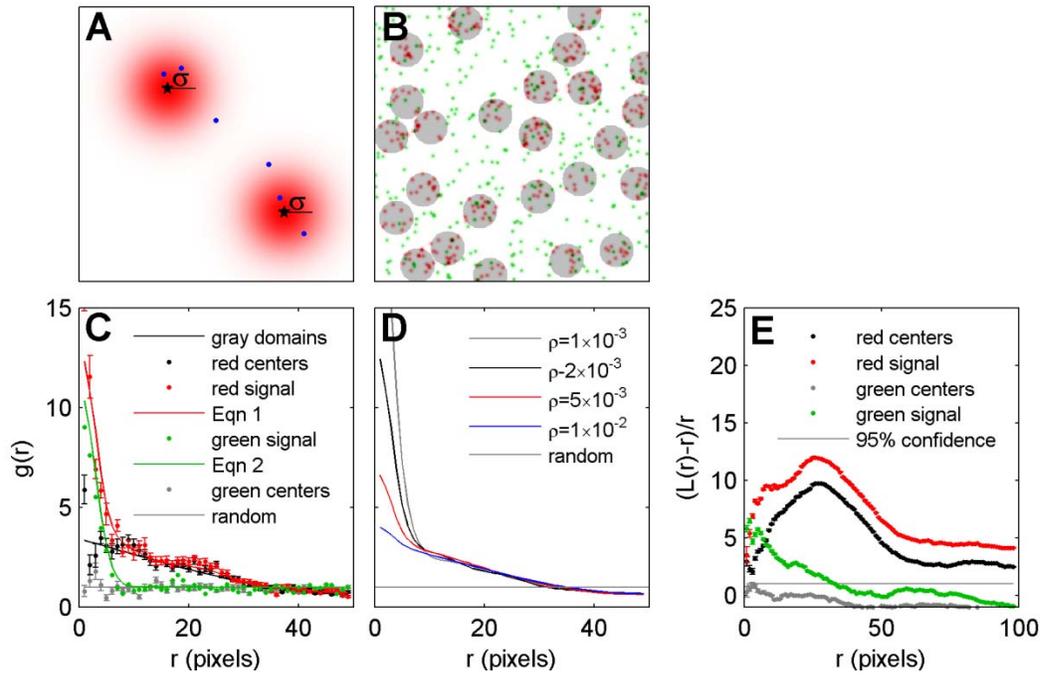

**Figure 1: Simulated demonstration of apparent clustering arising from over-counting individual labeled molecules with a finite PSF.**

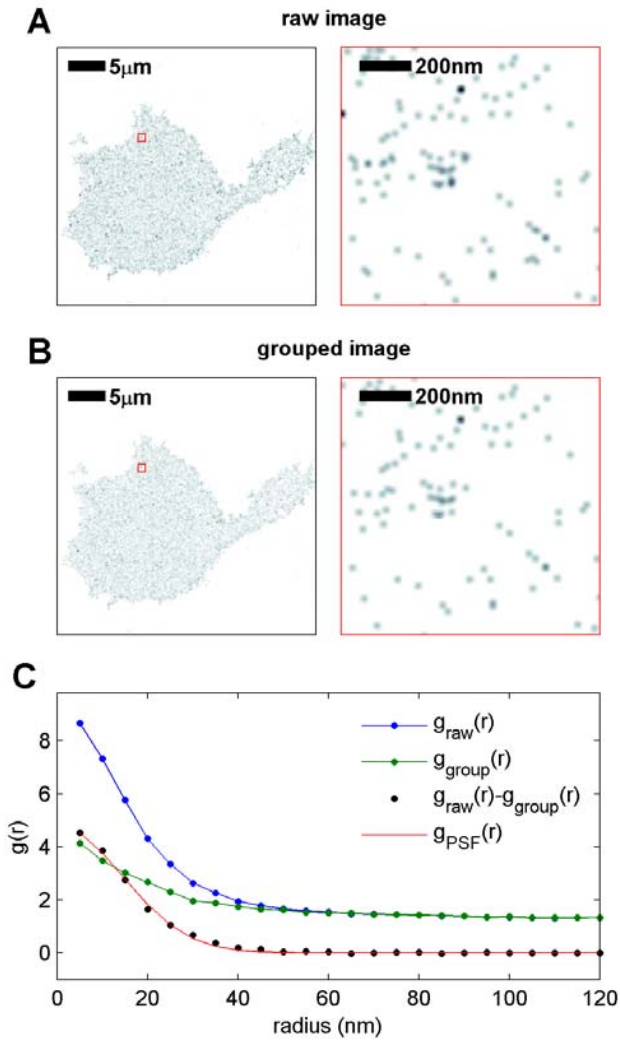

**Figure 2: Measuring effective resolution of reconstructed super-resolution images with explicit over-counting.**

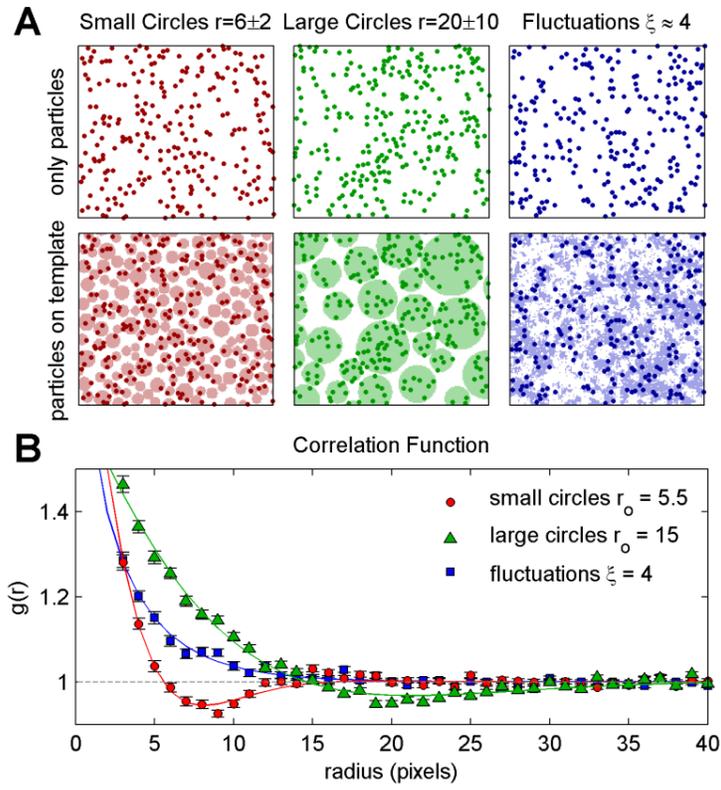

**Figure 3: Correlation functions quantify heterogeneity.**

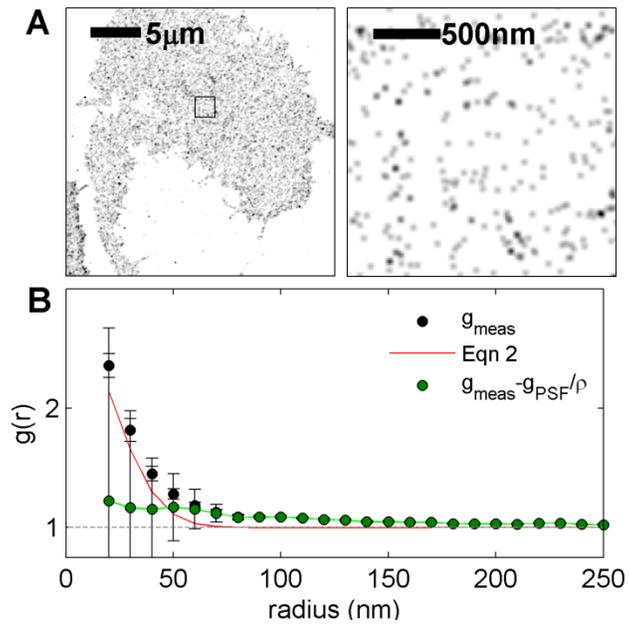

**Figure 4: Apparent clustering of IgE-FcεRI observed by STORM is well described as over-counting of multiple signals from individual labeled molecules.**

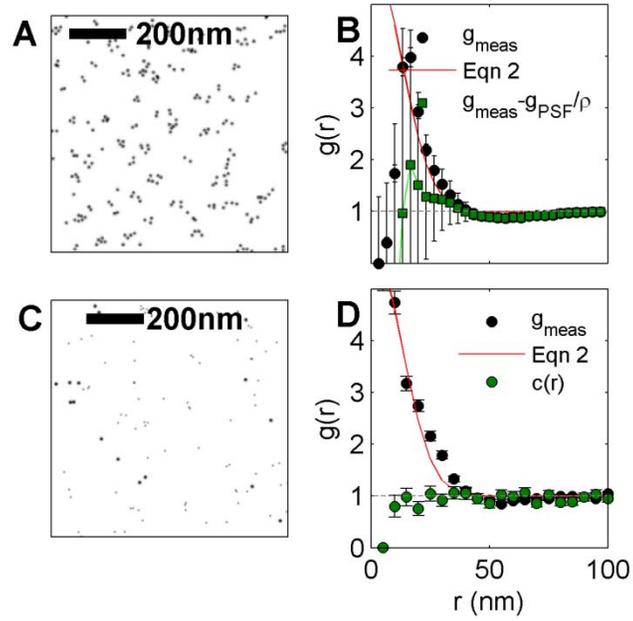

**Figure 5:** Apparent clustering of IgE-FcεRI observed by SEM is consistent with over-counting of multiple signals from individual labeled molecules.